# Topologías en el Internet de las Cosas Médicas (IoMT), revisión bibliográfica

# Topologies in the Internet of Medical Things (IoMT), literature review

Wilson Chango[1] https://orcid.org/0000-0003-3231-0153,
Teresa Olivares[2] https://orcid.org/0000-0001-9512-2745, Francisco Delicado[2] https://orcid.org/0000-0002-2150-7797

[1]*Escuela Superior Politécnica de Chimborazo*, Riobamba, Ecuador
wilson.chango@espoch.edu.ec

[2]*Universidad de Castilla – La Mancha*, Ciudad Real, España
teresa.olivares@uclm.es, francisco.delicado@uclm.es

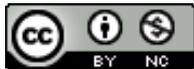

**Resumen**

La revisión bibliográfica es una fase fundamental en un proyecto de investigación, y debe garantizar la obtención de la información más relevante en el campo de estudio. El objetivo principal de este proyecto es conocer los trabajos relacionados con el Internet de las Cosas Médicas, en adelante (IoMT). Se analiza un total de 535 artículos buscados en Association for Computing Machinery, en Adelante ACM, Web of Science y Scopus el dominio de búsqueda fue IoMT. Se establecieron tres parámetros, (problemática, artefacto y evaluación del artefacto), esto de acuerdo a la Investigación de la Ciencia del Diseño, en Adelante DSR, es un enfoque de investigación para la construcción de artefactos para proporcionar una solución útil a un problema en cada dominio. La ecuación (Internet de las cosas Y malla) dio como resultado 535, (Internet de las cosas Y medicina) un total de 417 y finalmente (Internet de las cosas Y malla médica) con 8, esto significa que hay mucho por indagar en este dominio de investigación. Las ventajas identificadas en este tipo de topología es llevar los mensajes de un nodo a otro por diferentes caminos, no puede haber absolutamente ninguna interrupción en las comunicaciones, cada servidor tiene sus propias comunicaciones con todos los demás servidores. Los grandes datos procedentes de los dispositivos IoMT han influido drásticamente en las cuestiones de salud e informática. En este documento, se realiza una revisión de la literatura científica y se mapean las tendencias de investigación sobre el paradigma IoMT en el ámbito de la salud. Por último, este documento amplía la literatura, y los resultados de este estudio pueden servir de base para futuras investigaciones.

*Palabras clave:* IoMT, LPWAN, revisión bibliográfica, topologías de red, configuraciones de malla.

**Abstract**

The bibliographic review is a fundamental phase in a research project, and it must guarantee that the most relevant information in the field of study is obtained. Our main objective was to know the works related to the Internet of medical things, from now on (IoMT).  We analyzed a total of 535 articles searched in Association for Computing Machinery in Adelante ACM, Web of Science and Scopus the search domain was IoMT, we established 3 parameters, (problematic, artifact and artifact evaluation), this according to the Research of Design Science in Adelante DSR, is a research approach for the construction of artifacts to provide a useful solution to a problem in each domain. The equation (Internet of things AND mesh) resulted in 535, (Internet of things AND medicine) a total of 417 and finally (Internet of medical things AND mesh) with 8, this means that there is a lot to investigate in this research domain. The advantages identified in this type of topology is to carry messages from one node to another by different paths, there can be absolutely no interruption in communications, each server has its own communica-tions with all other servers. Health and IT issues have been drastically influenced by the large data from IoMT devic-es. In this paper, we conducted a review of the scientific literature and mapped research trends on the IoMT paradigm in the health domain. Finally, this paper expands on the liter-ature, and the findings of this study can serve as a basis for future studies.

**Introducción**

El Internet de las Cosas Médicas (IoMT), capta o analiza datos y los envía a otros dispositivos donde los usuarios finales son el personal sanitario. Esta información es útil para la toma de decisiones por parte de estas unidades médicas. Las propuestas para implementar este sistema utilizando las redes LPWAN como medio de comunicación aportan importantes ventajas sobre otras similares en cuanto a coste, cobertura y fácil adaptabilidad. Pero, ¿son las actuales topologías de red utilizadas para transmitir esta información las mejores opciones para el IoMT? Este artículo trata de responder a esta importante pregunta de investigación.

El IoMTha introducido un cambio revolucionario al facilitar la gestión de las enfermedades, mejorar los métodos de diagnóstico y tratamiento de las mismas y reducir los costes y los errores de la asistencia sanitaria. Este cambio ha tenido un gran impacto en la calidad de la asistencia sanitaria tanto para los pacientes como para el personal sanitario de primera línea. El IoMT es una combinación de dispositivos y aplicaciones médicas que se conectan a través de redes.

Muchos proveedores de servicios sanitarios están aprovechando las últimas tendencias de TI, como la virtualización, la nube, la movilidad y el análisis de grandes datos, para sentar las bases de las instalaciones de próxima generación. La mayoría de los proveedores están considerando cuidadosamente cómo las comunicaciones de datos, específicamente la red, pueden permitir la movilidad de los cuidadores, conectar el ecosistema de médicos, pacientes y dispositivos médicos, proporcionar una plataforma escalable para nuevos modelos de atención y, en última instancia, salvaguardar los datos de los pacientes. Los proveedores de servicios sanitarios deben tener en cuenta cinco requisitos esenciales para cumplir la promesa de prestar una atención continua: Ofrecer rendimiento, flexibilidad, seguridad, sencillez y economía (Performance, 2018).

Las LPWAN, son redes de largo alcance, alta cobertura y bajo consumo de energía. Esto se consigue realizando pequeñas transmisiones de datos con poco ancho de banda, ahorrando energía para una mayor penetración. Algunos ejemplos de estas redes son: Narrowband IoT (NB-IoT), Sigfox o LoRaWAN, una especificación de red creada por la LoRa Alliance. Sigfox y NB-IoT utilizan bandas con licencia, de ahí su coste de uso. Mientras que LoRa utiliza la banda libre, por lo que no tiene coste de uso. LoRaWAN tiene un bajo consumo de energía, un bajo ancho de banda y está pensada para un tráfico casi exclusivamente ascendente y una topología en estrella.

El principal interés de esta investigación consiste mejorar las redes y garantizar que se cumplan los objetivos de rendimiento (disponibilidad de conexión, cobertura global y, en ocasiones, en tiempo real), flexibilidad, seguridad, simplicidad y economía (bajo consumo de batería para los sensores). Este problema se llevará a cabo con la verificación de la comunicación entre dispositivos utilizando una topología adecuada, esto permite que los datos lleguen sin ningún problema a las unidades de salud para su respectivo análisis y toma de decisiones. Como se ha mencionado anteriormente, la topología utilizada en LoRaWAN es una topología en estrella. Creemos que sería bueno ampliar esta topología en estrella en determinados casos para aumentar principalmente la cobertura y evitar problemas de saturación y colisión en la pasarela de destino. Por lo tanto, este trabajo tiene como objetivo hacer un estudio de las topologías y soluciones de comunicación más utilizadas en el IoMT, analizar sus pros y sus contras e identificar aquellos aspectos de la comunicación del IoMT que requieren un mayor desarrollo e investigación para cumplir con los requisitos impuestos por el IoMT.

En este trabajo, por lo tanto, se realiza una revisión de la literatura científica y se mapea las tendencias de investigación sobre el paradigma de IoMT en el dominio de la salud, centrándose en las topologías, de las tres ecuaciones "Internet de las cosas Y malla", "Internet de las cosas Y medicina" e "Internet de las cosas Y malla médica" la última obtuvo un valor mínimo en comparación con las otras, esto significa que hay mucho por investigar en este dominio de investigación. Las ventajas identificadas en este tipo de topología es llevar los mensajes de un nodo a otro por diferentes caminos, no puede haber absolutamente ninguna interrupción en las comunicaciones, cada servidor tiene sus propias comunicaciones con todos los demás servidores.

En este trabajo se analizan las topologías más utilizadas en IoMT. El esquema del trabajo es el siguiente: En la sección 2 se presenta la metodología utilizada para el análisis bibliográfico aplicando criterios de búsqueda, en la sección 3 se describen las investigaciones realizadas en IoT e IoMT según la clasificación especificada en la sección 2, mientras que en la sección 4 se presentan los objetivos del trabajo en referencia a las topologías utilizadas en IoMT y, por último, en la sección 5 se presentan las conclusiones obtenidas del trabajo.

## Metodología

Para abordar el objetivo general identificado anteriormente, se revisaron 570 artículos científicos de la web del sitio y la base de datos Scopus, las palabras clave que se utilizaron para encontrar estos artículos fueron IoT, 'IoMT', 'Mesh Topology', en un rango de fechas de 2016-2021. Los artículos de investigación se clasificaron en tres grupos generales y se resumieron según el problema, el artefacto y la evolución del artefacto. La ecuación (Internet de las cosas Y malla) dio como resultado 276, (Internet de las cosas Y medicina) un total de 286 y finalmente (Internet de las cosas médicas Y malla) con solo 8. Esto significa que todavía queda mucho trabajo por hacer en las comunicaciones de IoMT y de redes de malla.

Para facilitar el proceso de búsqueda, las palabras clave debían estar presentes al menos en el título y en el resumen, además, los artículos debían estar publicados en inglés, se excluyeron los artículos que no estaban dentro del rango de fechas establecido, los que se encontraron duplicados en WoS, Scopus y los que no pertenecían a la informática, obteniendo un total de 570 artículos para analizar y, finalmente, seleccionar los 70 más relevantes.

**Figura 1**

*Esquema de la revisión del artículo*

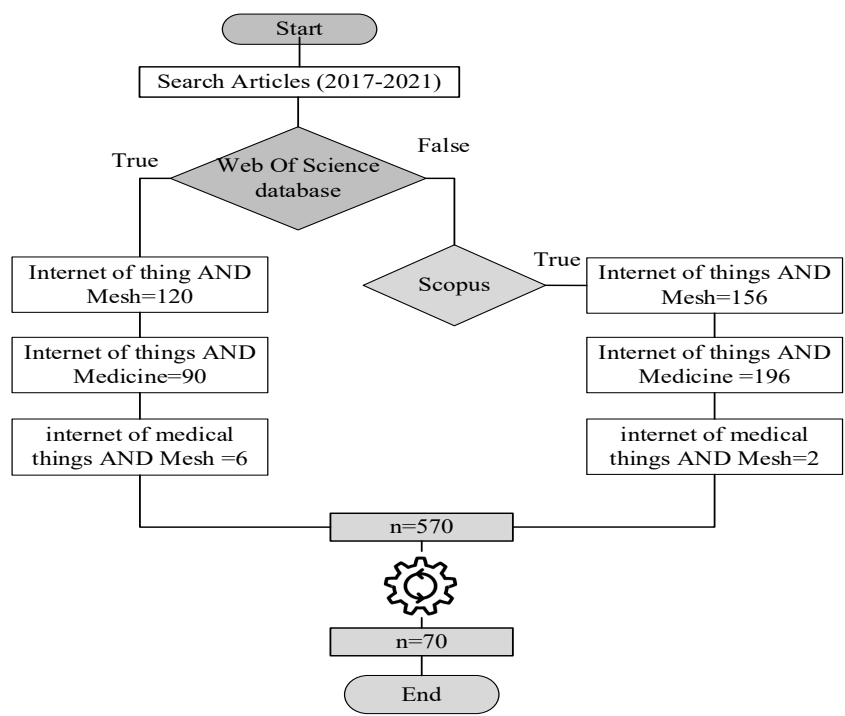

**Organización y discusión de la información encontrada**

Journal Citation Reports (JCR) cubre las publicaciones revisadas por pares más citadas del mundo y permite buscar el factor de impacto de una revista o grupo de revistas concreto y hacer comparaciones entre ellas. Cada grupo temático de revistas se divide en cuatro cuartiles: Q1, Q2, Q3, Q4, que corresponden respectivamente al grupo conformado por el primer 25% de las revistas del listado, el grupo que ocupa del 25 al 50%, el que se posiciona entre el 50 y el 75% y, por último, el que se ubica entre el 75 y el 100% del ranking ordenado. se confirma por aquellas que ocupan el primer cuartil Q1.

Posteriormente, mediante la lectura de los títulos, se excluyó cualquier artículo que no estuviera claramente relacionado con la medicina y el uso de cualquier topología, quedando una lista de 70 artículos, mientras que el resto se consideró de poca relevancia. La relevancia se refería principalmente a la conexión del artículo con el tema estudiado y a las publicaciones revisadas por pares más citadas en el mundo que permiten buscar el factor de impacto de una revista o grupo de revistas en particular y hacer comparaciones entre ellas. Cada grupo temático de revistas se divide en cuatro cuartiles (Q1, Q2, Q3, Q4), cuatro topologías de red (Start, Peer-to-Peer, Mesh, Hybrid) y el protocolo de comunicación utilizado en cada una de las investigaciones (ver Figura 2).

**Figura 2**

*Organización de la información*

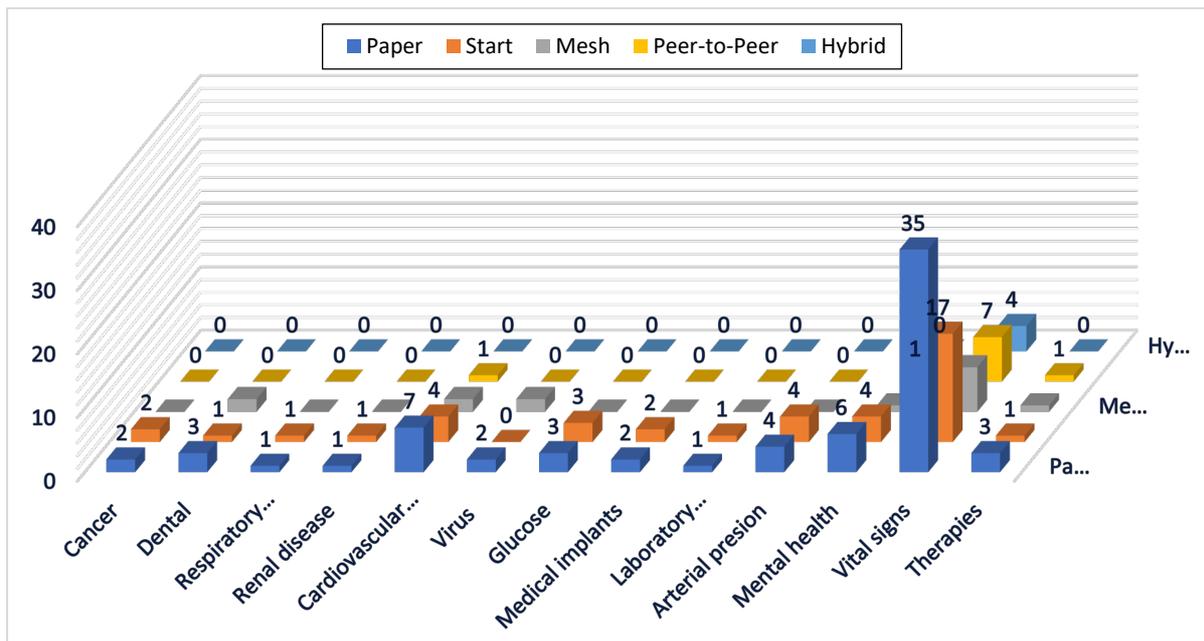

La Figura 2 muestra las investigaciones realizadas en el ámbito de la medicina (Cáncer Dental, Sistema Respiratorio, Enfermedades Renales, Enfermedades Cardiovasculares, Enfermedades Digestivas, Virus, Glucosa, Implantes Médicos, Monitorización de Laboratorio, Presión Arterial, Salud Mental, Signos Vitales, Signos Vitales -Seguridad, Terapias) frente a la topología de red

**Análisis de los resultados de la investigación**
**Figura 3**

*Aplicación principal en un Internet de las Cosas Médicas (IoMT).*

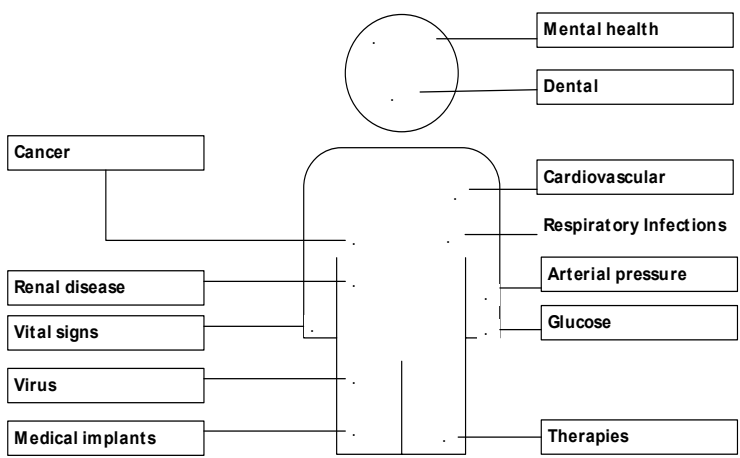

En la Tabla 1 se puede observar las aplicaciones en medicina con sus respectivas topologías. En cuanto al modelo de estrella se tiene un total de 41 de 70, esto significa que su uso es mayor; sin embargo, esta topología se basa en la centralización del cómputo, almacenamiento y control.

**Tabla 1**

*Clasificación de los trabajos en grupos de enfermedades y topologías*

| Medicine | Estrella | Malla | Hibrido |
|---|---|---|---|
| Cáncer | (Palani & Venkatalakshmi, 2019),(Khan et al., 2019) | | |
| Dental | (Liu et al., 2020) | (Alarifi et al., 2019),(Vellappally et al., 2019) | |
| Sistema Respiratorio | (Haoyu et al., 2019) | | |
| Enfermedad Renal | (Arulanthu & Perumal, 2020) | | |
| Cardiovascular | (Huang et al., 2019),(Al-Kaisey et al., 2020),(Pirbhulal et al., 2019) (Cubillos-Calvachi et al., 2020) | (Karthick & Manikandan, 2019),(Kan et al., 2015) | |
| Virus | | (Rani et al., 2019),(Song et al., 2018) | |
| Glucosa | (Leahy, 2008),(Gupta et al., 2019),(Cappon et al., 2017) | | |
| Implantes Médicos | (Santagati et al., 2020) | | |
| Laboratorio | (Kang et al., 2018) | | |
| Presión arterial | (Sood & Mahajan, 2019),(da Silva et al., 2019),(Farahani et al., 2020) (Sharman et al., 2020) | | |
| Salud Mental | (Sayeed et al., 2019),(Zilani et al., 2020),(Yadav et al., 2019) (Rachakonda et al., 2020),(Sarmento et al., 2020),(Rachakonda et al., 2019) | | |
| Signos vitales | (Awan et al., 2019),(Zanjal & Talmale, 2016),(Díaz de León-Castañeda, 2019) (Xing et al., 2018),(Evangeline & Lenin, 2019),(Ullah et al., 2017) (Hedrick et al., 2020),(Ng et al., 2020),(Rajasekaran et al., 2019) (Sánchez et al., 2019),(Mavrogiorgou et al., 2019),(Han et al., 2020) (Ignacio et al., 2019),(Qureshi & Krishnan, 2018),(Morzy et al., 2013) (Chen et al., 2020),(Nørfeldt et al., 2019) | (Kodali et al., 2016),(Cui et al., 2020)(Peng et al., 2017),(Plageras et al., 2016)(Medeiros et al., 2019),(Rajput & Brahimi, 2019a) (Rajput & Brahimi, 2019b),(Silvestre-Blanes et al., 2020) | (Kang et al., 2018),(Quincozes et al., 2019)(Garbhapu & Gopalan, 2017),(Choi et al., 2019) |
| Terapias | (Lam et al., 2020) | (Yu, 2020) | |

En cuanto a las redes Peer-to-Peer, se tiene un valor de 10 y su estructura consiste en que todos los nodos de la red están conectados entre sí por un enlace. De este modo, los nodos de una topología Peer-to-Peer están conectados punto a punto con los demás nodos de la red, mientras que los dispositivos finales están conectados al nodo de red más cercano. Así, este tipo de configuración se adapta a la situación en la que se encuentra la red con respecto al tráfico de información. Por último, las redes Mesh, un total de 15, se componen básicamente de tres clases de nodos: gateways o Dispositivos de Funcionalidad Completa (FFD), sensores/actuadores-routers o Dispositivos de Funcionalidad Reducida (RFD).

Por último, la topología híbrida con un valor de 4 se deriva de la unión de dos topologías de red (topología estrella-autobús topología estrella-anillo). Su implementación es debido a la complejidad de la solución de red y tiene un costo muy alto debido a su administración y mantenimiento. También proporciona los beneficios que ofrecen los otros diseños. Esto significa que puede tener la facilidad de localizar problemas de la organización en estrella o las facilidades económicas que ofrece la red de bus. A continuación, se describen de forma general las investigaciones realizadas en cada área de la medicina, considerando las topologías descritas en el apartado anterior.

**Tabla 2**

*Clasificación de los trabajos en grupos de enfermedades y tecnologías*

| Medicine | Zigbee | WIFI | 3G/4G/5G | LoraWan |
|---|---|---|---|---|
| Cáncer | | (Palani & Venkatalakshmi, 2019),(Khan et al., 2019) | | |
| Dental | (Alarifi et al., 2019) | | (Liu et al., 2020) | (Vellappally et al., 2019) |
| Sistema Respiratorio | | (Haoyu et al., 2019) | | |
| Enfermedad Renal | | | (Arulanthu & Perumal, 2020) | |
| Cardiovascular | | (Huang et al., 2019),(Al-Kaisey et al., 2020),(Pirbhulal et al., 2019) (Cubillos-Calvachi et al., 2020),(Ali et al., 2020),(Karthick & Manikandan, 2019) | (Kan et al., 2015) | |
| Virus | | | (Rani et al., 2019),(Song et al., 2018) | |
| Glucosa | (Leahy, 2008) | (Leahy, 2008),(Gupta et al., 2019) | (Cappon et al., 2017) | |
| Implantes Médicos | | (Mumtaz et al., 2018) | | |
| Laboratorio | | (Kang et al., 2018) | | |
| Presión arterial | | (Sood & Mahajan, 2019),(da Silva et al., 2019),(Farahani et al., 2020) (Sharman et al., 2020) | | |
| Salud Mental | (Sayeed et al., 2019),(Rachakonda et al., 2020) | (Rachakonda et al., 2020) | (Yadav et al., 2019),(Sarmento et al., 2020) | |
| Signos vitales | (Kodali et al., 2016),(Ullah et al., 2017),(Hedrick et al., 2020) (Peng et al., 2017),(Rajasekaran et al., 2019),(Kang et al., 2018) (Garbhapu & Gopalan, 2017),(Elsts et al., 2018),(Latif et al., 2020) (Mavrogiorgou et al., 2019),(Rajput & Brahimi, 2019a),(Rajput & Brahimi, 2019b) (Han et al., 2020),(Silvestre-Blanes et al., 2020) | (Zanjal & Talmale, 2016),(Díaz de León-Castañeda, 2019),(Xing et al., 2018) (Evangeline & Lenin, 2019),(Ullah et al., 2017),(Quincozes et al., 2019) (Plageras et al., 2016),(Medeiros et al., 2019),(Sánchez et al., 2019) (Ignacio et al., 2019),(Qureshi & Krishnan, 2018),(Toor et al., 2020) (Rubí & Gondim, 2019),(Choi et al., 2019),(Morzy et al., 2013) (Chen et al., 2020),(Nørfeldt et al., 2019) | (Cui et al., 2020),(Zanjal & Talmale, 2016) | |
| Terapias | | (Yu, 2020),(Lam et al., 2020),(Sanders et al., 2019) | | |

**El cáncer**

Es una enfermedad grave que ha matado a muchas personas durante los últimos años debido al hábito alimenticio de los seres humanos. En 2019, (Palani & Venkatalakshmi, 2019) proponen un nuevo modelado predictivo basado en el Internet de las Cosas (IoT) mediante el uso del aumento de clústeres difusos y la clasificación para predecir la enfermedad del cáncer de pulmón; sin embargo, no se discute qué tipo de topologías, protocolos y tecnologías utiliza, pero propone una arquitectura, utiliza una topología de estrella como diseño de red y para la comunicación de dispositivos IoT se realiza por WIFI.

Los dispositivos basados en WIFI necesitan una buena batería de respaldo si uno quiere utilizarlos por más de 10 horas aproximadamente, una sola red basada en WIFI puede tener un tamaño de red de hasta 2007 nodos, el WIFI ha sido estandarizado de acuerdo al estándar IEEE 802.11.x. Existen varias versiones del protocolo donde la x se sustituye por a, b, g, n, etc. que son diferentes versiones de WiFi, utilizadas para redes de área PAN y WLAN con un alcance medio de 30 a 100 metros, (Khan et al., 2019) propone una infraestructura, en la que los mensajes no se envían directamente, es decir, tienen que pasar por un nodo de acceso (módem/routers), lo que podemos concluir que es una topología en estrella y utiliza la tecnología WIFI.

**Dental**

En 2020, (Liu et al., 2020) proponen un sistema inteligente de salud dental-IoT basado en hardware inteligente, aprendizaje profundo y terminal móvil, con el objetivo de explorar la viabilidad de su aplicación en el cuidado de la salud dental en el hogar.

La arquitectura de red del sistema iHome dental health-IoT consta de tres capas de red: 1) capa de servicio médico dental; 2) capa de servicio dental inteligente; 3) capa de adquisición de datos de imágenes dentales. Los datos de las imágenes dentales se cargan en la capa de servicio dental inteligente a través de la red (Wi-Fi, 3G/4G). A continuación, (Alarifi et al., 2019) examinan continuamente los patrones faciales, las patologías y la discrepancia cefalométrica para tomar la decisión sobre el proceso de extracción y no extracción de dientes. Toda la información relacionada con los dientes se transmite a través de la señal de radiofrecuencia, las redes Zigbee transmiten en radiofrecuencia, esto implica que si se utiliza Zigbee como Z-Wave vamos a requerir un hub, puente o concentrador que será el punto del sistema que se conecta a internet. Esta señal Wi-Fi se compartirá entre el resto de dispositivos de la red sin necesidad de que cada uno de ellos se conecte al router de forma individual. Tanto Zigbee como Z-Wave funcionan a través de una red de malla.

**Enfermedades respiratorias**

Está relacionado con el sueño. La prueba de referencia para el diagnóstico es la polisomnografía, que requiere mucho tiempo y es cara. (Haoyu et al., 2019), propone la siguiente arquitectura, todos los datos pasan por un nodo o Gateway esto significa que en esta investigación se utiliza la topología de estrella junto con la tecnología WIFI. La parte principal de esta sección es el sensor de pulso y proximidad MAX3010X conectado al ESP32 que puede servir como un módulo de transmisión de datos portátil a través de Wi-Fi. Para la comunicación se establece una frecuencia de muestreo de 60 Hz con una resolución de 10 bits. Una conexión en serie a través de Wi-Fi se utiliza para enviar estas señales al módulo de remodelación y detección de errores, las ventajas y desventajas de esta tecnología se describieron en la sección de cáncer.

**Enfermedad renal**

(Arulanthu & Perumal, 2020), presentan un sistema de apoyo a las decisiones médicas en línea (OMDSS) para la predicción de la enfermedad renal crónica (ERC). El modelo presentado incluye un conjunto de etapas, a saber, la recopilación de datos, el preprocesamiento y la clasificación de datos médicos para la predicción de la enfermedad renal crónica, habiendo dicho que se propone un nuevo OMDSS para la predicción de la ERC y ofrece servicios sanitarios eficaces a los pacientes.

Los datos de los pacientes son recogidos por dispositivos IoT. En general, el sensor conectado al ser humano recoge datos regularmente en un intervalo de tiempo específico. El OMDSS presentado hace uso de la red 4G para transmitir los datos observados al CDS. Hay dos conceptos fundamentales que sustentan el éxito de las redes de telefonía 4G, uno es el modelo de banda ancha móvil y el otro es la convergencia de redes.

**Enfermedades cardiovasculares**

(Huang et al., 2019) presentan un sistema para la adquisición de datos de lípidos en sangre basado en un smartphone para controlar el nivel de lípidos en sangre, la propuesta de adquisición de datos fotoquímicos de lípidos en sangre basada en un smartphone con la tira reactiva y la arquitectura IoMT para la gestión de los lípidos en sangre. Esta información se envía al smartphone a través del cable OTG. Esta parte puede considerarse como la capa de recogida de parámetros de la función de lípidos en sangre en la arquitectura IoMT. Los datos serán finalmente subidos a la nube para su cálculo y almacenamiento los pacientes y los médicos pueden acceder a los datos en la nube desde sus respectivos terminales a través de la conexión WIFI.

(Al-Kaisey et al., 2020) también utilizan para la comunicación un nodo de entrada para el latido de las señales de ECG, también utiliza una topología en estrella, por su parte (Pirbhulal et al., 2019) proponen una arquitectura para resolver un problema crítico en la implementación de la seguridad para la transmisión de la información de salud, esto permite proporcionar la privacidad de los datos y la validación de la información de un paciente sobre el entorno de la red de manera eficiente en el uso de los recursos.

**Virus**

El mosquito es uno de los insectos fatales que sopla varios patógenos como el Chikungunya, que es una enfermedad instintiva que se propaga rápidamente en varias partes del país, por lo tanto, hay una necesidad de medidas preventivas para esta enfermedad, los nodos sensores y los teléfonos celulares se utilizan para detectar y recopilar datos para la comunicación utilizando 4G y una topología de malla (Rani et al., 2019)**.** Estos datos son procesados antes de ser transmitidos a la nube.

Dicho lo anterior, (Song et al., 2018) utilizan una topología de malla y diseñaron una plataforma de cribado móvil basada en teléfonos inteligentes, sencilla y barata, denominada "smart connected cup" (SCC), para realizar diagnósticos moleculares rápidos, conectados y cuantitativos. Esta plataforma combina el ensayo bioluminiscente en tiempo real y la amplificación isotérmica mediada por bucle (BART-LAMP) con la detección basada en el teléfono inteligente, la transmisión de los resultados de las pruebas al registro del paciente y al consultorio del médico; y la comunicación de las pruebas se realiza con el Sistema de Posicionamiento Global (GPS) y 4G, el uso de teléfonos inteligentes mejora las capacidades más allá de lo que está disponible con los instrumentos existentes.

**Glucosa**

El suministro de insulina a través de sensores a personas con diabetes en un reto para el IoT, la terapia de reemplazo celular más prometedora para niños con diabetes tipo 1 es un páncreas artificial de bucle cerrado que incorpora sensores de glucosa continuos y bombas de insulina, (Leahy, 2008) combina una bomba externa y un sensor con un algoritmo de tasa de infusión de insulina variable diseñado para emular las características fisiológicas de la célula, para la comunicación utiliza tecnología WIFI y topología en estrella, (Cappon et al., 2017) realizan la conexión de un sensor - App móvil - Transmisor inteligente para la monitorización de la glucosa en personas con diabetes, los resultados muestran una mejora estadísticamente significativa de todas las métricas glucémicas.

Se puede observar un sensor de aguja mínimamente invasivo, generalmente insertado en el tejido subcutáneo, el abdomen o el brazo, que mide una señal de corriente eléctrica generada por la reacción de la glucosa oxidasa. Esta señal es proporcional a la concentración de glucosa disponible en el líquido intersticial, que se convierte en una concentración de glucosa mediante un procedimiento de calibración que suele realizarse dos veces al día. Los dispositivos se conectan mediante WIFI para enviar la información.

**Implantes médicos**

El mayor obstáculo para los implantes en red es la naturaleza dieléctrica del cuerpo humano, (Santagati et al., 2020) han demostrado que se pueden generar y recibir ondas ultrasónicas de forma eficiente con componentes milimétricos de baja potencia, y que a pesar de la pérdida de conversión que introducen los transductores ultrasónicos, a partir de esta investigación fundamental, construyó un prototipo que puede ser la base para construir futuros implantes médicos y dispositivos vestibles.

Por último, (Mumtaz et al., 2018) se dieron cuenta de que el uso de un enfoque de estimación de la transmisión para mejorar la precisión de la fiabilidad de la transmisión y la estimación de acuerdo con la dinámica del sistema, luego se optimizó mediante la formulación de un problema de minimización restringido.

**Signos vitales**

Se está investigando mucho para reducir el costo y aumentar la eficiencia en la industria médica, (Awan et al., 2019) diseñan un protocolo de enrutamiento entre los nodos de sensores de tal manera que tiene un retraso mínimo y un mayor rendimiento para los paquetes de emergencia utilizando IoT y un consumo de energía óptimo para una mayor vida útil de la red, así como la utilización eficiente de los recursos escasos. El rendimiento del protocolo propuesto se evalúa con dos técnicas de enrutamiento del estado de la técnica iM-SIMPLE y enrutamiento optimizado rentable y eficiente de la energía, esto es crucial para el seguimiento de la atención de la salud de los pacientes que no pueden tener el acceso rápido a un hospital. Los datos fluyen a través de un nodo de acceso (routers) y los dispositivos se comunican utilizando el WIFI.

**Terapias**

La lentitud, la falta de fiabilidad y la desorganización en el sistema de diagnóstico médico de lesiones deportivas, (Yu, 2020) lo mejora mediante una aplicación para el diagnóstico móvil y la gestión de datos del usuario. En primer lugar, la capa de aplicación, la capa de red y la estructura de la capa de percepción IoT se utilizan para planificar la jerarquía general del sistema médico de lesiones deportivas, y luego el sistema se divide en tres módulos, que son el módulo de adquisición de parámetros fisiológicos del usuario, el módulo de procesamiento de parámetros fisiológicos del usuario y el módulo de diagnóstico de daños por movimiento, Por último, los sensores (red de malla), los servidores, los protocolos de comunicación específicos, representan un cambio inteligente hacia procesos de fabricación más interconectados en los que las entidades individuales dentro de la cadena de suministro se comunican entre sí para lograr una mayor flexibilidad y capacidad de respuesta en la fabricación general y una fabricación más eficiente para reducir el costo de producción (Lam et al., 2020), esto debido a que un gran número de personas deben utilizar prótesis (Sanders et al., 2019).

**Salud mental**

La epilepsia se caracteriza por la recurrencia de convulsiones espontáneas y tiene un impacto negativo considerable en la calidad y la esperanza de vida del paciente. La tecnología UWB solo transmite en distancias cortas (hasta 10 metros), pero tiene la ventaja de lograr un ancho de banda muy elevado (hasta 480 Mbps), consumiendo poca energía. Es ideal para la transferencia inalámbrica de contenidos multimedia de alta calidad, como vídeos, entre dispositivos electrónicos de consumo y periféricos informáticos (Sayeed et al., 2019). Además, los métodos de evaluación de la Ataxia son engorrosos y no permiten un control y seguimiento regular de los pacientes. Una de las tareas más difíciles es detectar las diferentes anomalías de la marcha en los pacientes con Ataxia (Zilani et al., 2020). El estrés crónico, el consumo de alimentos incontrolado o no supervisado y la obesidad están estrechamente relacionados, incluso implicando ciertas adaptaciones neurológicas (Rachakonda et al., 2020). En conclusión, los accidentes cerebrovasculares se encuentran entre las tres principales causas de muerte en todo el mundo. Además, los accidentes cerebrovasculares son una de las principales causas de morbilidad, hospitalizaciones y discapacidades adquiridas (Sarmento et al., 2020).

**Presión arterial**

La hipertensión es una enfermedad crónica que provoca riesgo de diferentes tipos de trastornos, como ataques de hipertensión, accidentes cerebrovasculares, insuficiencia renal y enfermedades cardiovasculares (Sood & Mahajan, 2019), lo más importante es controlar las prescripciones designadas y revisar constantemente la presión arterial para evitar el conocido "efecto bata blanca" (da Silva et al., 2019). Del mismo modo, la relación entre la tecnología y la atención sanitaria, debido al auge del Internet de las cosas inteligentes (IoT), la inteligencia artificial (IA) y la rápida adopción pública de wearables de grado médico, se ha transformado drásticamente en los últimos años. En el mismo contexto, (Farahani et al., 2020) proponen una arquitectura holística de IoT eHealth impulsada por la IA y basada en el concepto de aprendizaje automático colaborativo, en la que la inteligencia se distribuye a través de la capa del dispositivo, la capa del borde/la niebla y la capa de la nube, actualmente hay más de 3. 000 dispositivos de AP disponibles en el mercado, pero muchos de ellos no tienen publicados datos de pruebas de precisión según los estándares científicos establecidos (Sharman et al., 2020).

**Laboratorio de monitorización**

Las operaciones ininterrumpidas y la continuidad del negocio son requisitos clave para cualquier edificio altamente automatizado situado bajo el paradigma de la industria 4.0, para lo cual la Calidad de la Energía juega un papel importante. (Alonso-Rosa et al., 2018) describen un novedoso sensor de bajo coste para el Internet de las Cosas que mide y analiza la calidad de la energía a la entrada de cualquier dispositivo de corriente alterna (CA), proporcionando un sistema de detección y análisis temprano que monitoriza aquellas variables críticas que varían dentro de la instalación y permite anticiparse a los fallos con alertas en fase temprana basadas en el procesamiento del flujo de datos.

### El Problema de La Topología De La Red

Existen tres modelos topológicos básicos para la IOT: el modelo en estrella, el modelo en malla y el modelo híbrido. Estructuralmente, el modelo en estrella se basa en la centralización de la computación, el almacenamiento y el control. En una red Star IoT, un dispositivo final solo está conectado a un único nodo de red, que actúa como nodo central para el dispositivo. A su vez, cada uno de estos nodos de red está conectado a un nodo servidor, que podría ser una conexión a otro nodo de red con capacidades superiores o funcionalidades diferentes, cumpliendo este último el papel de nodo central de mayor jerarquía (K. & Desai, 2016).

De esta forma, para que dos dispositivos finales se comuniquen entre sí, la información debe viajar hasta el nodo central que comunica las rutas a ambos dispositivos y luego completar el camino desde el nodo central hasta el destinatario. Este tipo de estructura de red tiene ventajas en un entorno IoT debido a su facilidad de configuración, refiriéndose a ella como añadir y eliminar dispositivos finales y detectar fallos. Además, esta configuración, normalmente cargada en un dispositivo central que supervisa la resolución de todas esas complejas situaciones, permite que el rendimiento de la red sea consistente, predecible y bueno: baja latencia y suficiente ancho de banda. La latencia, una preocupación para las aplicaciones de IoT, disminuiría con el uso de este tipo de topologías, ya que se reduce el número de saltos necesarios para transportar la información hasta el destino. A pesar de todo esto, las redes en estrella tienen serias desventajas que afectan al rendimiento en entornos IoT. La centralización de las funcionalidades resuelve muchos problemas, pero presenta una de las desventajas más graves: un único punto de fallo. Si el dispositivo central con el que se comunican los diferentes nodos de la red falla, el rendimiento de la red se ve completamente perjudicado. En cambio, si los nodos más cercanos a los dispositivos finales fallan -hablando de una red en estrella en forma jerárquica- el sector de la red que falla puede aislarse rápidamente, mientras el resto de la red funciona correctamente. Las redes en estrella también suelen tener dificultades con respecto a las interferencias de radio, el alcance de la transmisión (limitado al alcance de la transmisión del dispositivo que actúa como nodo central) y el consumo de energía, que depende de la distancia entre los dispositivos conectados, recordando que los dispositivos de la IO funcionan en su mayoría con baterías.

**Tabla 3**

*Topologías Usadas en IoT*

| Topologías | Resultado |
|---|---|
| Start | 41 |
| Peer-to-Peer | 10 |
| Mesh | 15 |
| Hybrid | 4 |
| Sum | 70 |

En cuanto a las redes Peer-to-Peer, tienen una estructura en la que todos los nodos de la red están conectados entre sí, es decir, existe un enlace permanente entre cada nodo de la red. De este modo, este tipo de configuración permite una conexión adaptativa a la situación en la que se encuentra la red con respecto al tráfico de información. Las redes peer-to-peer no son tan fáciles de configurar, ya que es necesario establecer los enlaces entre cada uno de los nodos de la red añadidos y los existentes, y luego conectar los dispositivos finales a ellos. En una topología de malla cada nodo de la red está conectado a al menos otros dos nodos, mientras que la topología híbrida se divide en topología estrella y topología anillo (Montiveros et al., 2018).

**Conclusiones**

En este artículo se presenta una metodología para realizar una revisión bibliográfica, como es la Ciencia del Diseño que es una metodología de investigación en tecnologías de la información basada en resultados analizando tres parámetros importantes como son el problema, el artefacto y la forma en que el artefacto fue evaluado, a través de una macro búsqueda que permite identificar los documentos relacionados con el tema de investigación. Las estrategias de búsqueda, organización y análisis de la información, permiten tanto la obtención de los documentos referentes a un tema de investigación, como su sistematización y estructuración para analizar las principales características del conjunto de documentos en estudio. Se presentó un caso sobre un tipo de aplicación de IoT, el tema se denomina "Internet de las cosas médicas (IoMT) un reto en la topología", donde los principales problemas son las topologías utilizadas.

**Referencias**